\documentclass[aps,amsfonts,twocolumn,nofootinbib]{revtex4-1}
\usepackage[utf8]{inputenc}
\usepackage{epsfig}
\usepackage[spanish]{babel}
\usepackage[utf8]{inputenc}
\usepackage{parskip}
\usepackage{graphicx}
\usepackage{slashed}
\usepackage{amsmath}
\usepackage{amsbsy}
\usepackage{bm}
\usepackage{color}
\usepackage{epsfig}
\usepackage{stackengine}
\stackMath
\newcommand{\ee}{\end{equation}}
\newcommand{\bb}{\begin{equation}}
\newcommand{\eqb}{\begin{eqnarray}}
\newcommand{\eqf}{\end{eqnarray}}

\newcommand{\1}{{\'{\i}}}

\def\1{\'{\i}}

\def\1{\'{\i}}

\begin{document}
\title{ Deformed Quantum Mechanics and the Landau Problem}
\author{J. Gamboa}
\email{jorge.gamboa@usach.cl}
\affiliation{Departmento de F\1sica, Universidad de Santiago de Chile, Casilla 307, Santiago, Chile}
\author{F. M\'endez}
\email{fernando.mendez@usach.cl}
\affiliation{Departmento de F\1sica, Universidad de Santiago de Chile, Casilla 307, Santiago, Chile}
\begin{abstract} 
 A deformation of the Landau problem  based on a modification of Fock algebra is considered. Systems with the Hamiltonians   $f ({\hat H})$ where ${\hat H}$ is the Landau Hamiltonian in the lowest level are discussed. The case $f({\hat H}) = \alpha_1 {\hat H} + \alpha_2 {\hat H}^2$ is  studied  and it is shown that in this particular example,  parameters of the problem can be fixed by using the quadratic Zeeman effect data and the Breit-Rabi formula.  {{ The proposed approach allows
to solve exactly   Landau-like families of problems not previously discussed in the literature.}} \end{abstract}
\maketitle

 In the last years the possibility of deforming quantum mechanics has been widely discussed 
 from different points of view as a way to find a solution to several fundamental 
 problems \cite{Banks:1983by,Hawking:1979ig,kibble,weinberg}. 

In this regard, issues such as states evolution,  algebras of observables
and unitarity  of the S-matrix have been re-analyzed and, as a consequence, new approaches and concepts 
in quantum theory have been introduced \cite{Banks:2020wul}.
 
It is expected that these  new concepts and approaches  might shed light on different problems, ranging from
the black hole physics ( for example the evaporation process \cite{susskind,polchinski})  to solid state physics 
 (such as an explanation to  the high temperature superconductivity \cite{putue}).

 
 Although a final answer to these questions has  not been reached yet, the  methods developed 
 in this field  bring different  insight and  perspectives and they also provide  useful calculation techniques to 
 tackle new and old problems. 
 Some of these calculations methods were developed almost at the same time as quantum mechanics.
 For example, the non-commutativity of spacetime as an ultraviolet regulator \cite{snyder,yang} or the phase space quantum mechanics \cite{saco}. Others, however, such as deformed commutator structures, have led to  the development of
 new and  important areas of mathematics such as non-commutative geometry \cite{connes}, quantum groups \cite{kulito}, and deformed Poissonian geometry \cite{konse}. 

In the present manuscript,  we use some of the results outlined before in order to explore a non-commutative system 
(in the sense of non-commutative quantum mechanics) together with a deformation of commutators (as those appearing 
in the study of  quantum algebras). We focus in a concrete example, namely the  Landau problem,  and its relation with  the quadratic Zeeman effect.

More precisely we will introduce a deformation in the ladder operators -- inspired by a modification of Fock  algebra -- for the Landau problem that emerges from a non-commutative quantum mechanical system 
 and then we will show how this modification might be  useful in the understanding the quadratic Zeeman effect and how the free deformation parameters can be set by experiments. 
 
 
 In the context of non-commutative quantum mechanics the Landau problem can be understood by defining  
 the operator $\aleph$  \cite{nos01}
 \bb
 \aleph=  \frac{\theta^2}{4} {\bf p}^2+x^2+y^2 -\theta(x\,p_y -y\,p_x), \label{a1}
 \ee 
 where $\{x,y,p_x,p_y\}$ satisfy the Heisenberg algebra and $\theta$ is a parameter with dimensions of  (energy)$^{-2}$.
The operator $\aleph$ formally describes  a charged two-dimensional harmonic oscillator  with mass $m=2/\theta^2$, frequency $\omega =\theta$, in an external magnetic field $B=\theta$\footnote{Through the paper we use natural units 
$c=1,\hbar=1$.}. 
 
 The eigenvalues of $\aleph$
have been  discussed  in the context of non-commutative quantum mechanics \cite{refers} and 
the important fact is that in a  two dimensional non-commutative space,  any  central field  $V(|{\bf x}|^2)$ becomes $V(\aleph)$ and therefore,  general problems in two dimension can be explicitly addressed.
  
 Let us  discuss this last fact. Consider the potential $V(|\hat{{\bf x}}|^2)$ in a space where coordinates
 $\{\hat{x},\hat{y}\}$ satisfy  $[\hat{x},\hat{y}]=\imath\,\theta$. By changing the basis of the algebra, this potential
 satisfies
%
%
%
 \begin{eqnarray}
\label{sny5}
V(\hat{x}^2+\hat{y}^2) &=& V\left( \frac{\theta^2}{4} {\bf p}^2+x^2+y^2 -\theta(x\,p_y -y\,p_x)\right) \nonumber
\\
& = & V(H_{\mbox{\tiny{HO}}} -\theta\,L_z)
\\
&\equiv& V(\aleph).
 \end{eqnarray}
The  variables $\{x,y,p_x,p_y\}$ satisfy the Heisenberg algebra,   
the operator  $H_{\mbox{\tiny{HO}}}$ denotes a two-dimensional harmonic oscillator as in (\ref{a1}), while the operator $L_z=(x\,p_y -y\,p_x)$ is a conserved quantity in the 
sense $\left[L_z, H_{\mbox{\tiny{HO}}}\right]=0$. 


This system is an example  of non-commutative quantum mechanics and the analysis can 
be extended to cases where the whole phase space is non-commutative -- by introducing momentum 
variables $\{p_x,p_y\}$ satisfying $[p_x,p_y]=i B$ with $B$ a constant -- giving rise to new interesting 
features such as a sort of phase transition for $\theta= \frac{1}{B}$.
  

The operator $\aleph$ can be diagonalized in terms of ladder operators. That is, by defining
\eqb 
&&a_{\pm}= \frac{1}{\sqrt{2}} \left(a_y\pm i a_x\right), \nonumber
\\
&&a^{\dagger}_{\pm} =\frac{1}{\sqrt{2}} \left(a^\dagger_y\mp i a^\dagger_x\right). \label{sny8}
\eqf
 with $a_x = (\theta)^{-\frac12}(x+i (\theta/2) p_x)$ and a similar definition for $a_y$,
 being $a^\dagger$ the conjugate transposed operator. The operators previously defined 
satisfy  the following algebra
\begin{equation}
\label{conmua}
[a_\pm,a_\pm^\dagger]=1,\quad [a_\pm,a_\mp]=0,
 \end{equation}
 and therefore, spaces $+$ and $-$ are orthogonal.
 
Then, $\aleph$ operator is diagonal in the base  $|n_-,n_+\rangle$, that is 
\bb 
\aleph |n_-,n_+\rangle = \Lambda_{n_- n_+}|n_-,n_+\rangle, 
\label{snyx2}
\ee
with   
\bb
\Lambda_{n_-n_+} = \theta \left(2n_- +1\right), \label{sny9}
\ee
thus,  this is an infinitely degenerate system.

In (\ref{sny9}) the  $\theta$  parameter emerges as an effective frequency  and 
the factor $1$ there is  the zero point energy which can be removed
by a normal order prescription of operators.
\smallskip
    
The total Hamiltonian for a particle in this potential  reads
\begin{equation}
\label{sny4}
 H= \frac{1}{2M} {\bf p}^2 + V(\aleph),
 \end{equation} 
where $M$ is a mass scale. It has been shown \cite{nos01} that this Hamiltonian
is equivalent to the Landau problem  in the lowest level -- in the strong magnetic field regime --  in the limit $M\to\infty$ 
for a linear potential $V(\aleph) =\Omega\aleph$ with $ \Omega$ a constant with 
dimensions of (energy)$^{3}$. 

Therefore, we  study the case in which $M\to \infty$ and the Hamiltonian turn out to be
\begin{equation}
\label{ourmodel}
H= V(\aleph).
\end{equation}

Since $\aleph$ satisfies (\ref{snyx2}) 
\bb
 H | n \rangle = V (\aleph) | n \rangle = V( \epsilon_n ) |n\rangle, 
\label{sny10}
 \ee  
where $n$  is a notation for the set of quantum numbers $\{n_+,n_-\}$ and 
$\epsilon_n$ denotes the eigenvalues of $\aleph$.

%
%
 For  $V(\aleph)$  expandable 
  in a Taylor series around a small parameter  $\lambda $   \cite{gross}
\bb
V(\aleph) = \aleph + \lambda \aleph^2 + \cdots, \label{sny11}
\ee  
an   {\it  effective Hamiltonian }  can be defined
 \bb
 H_{\mbox{\tiny{eff}}} =: V(\aleph)= \aleph + \lambda \aleph^2 + \cdots, \label{sny12} 
 \ee
 and then 
 \bb 
 H_{\mbox{\tiny{eff}}}|n\rangle = V(\epsilon_n ) |n\rangle. \label{sny13}
 \ee 
 
 An example of this system is the Euler-Heisenberg effective Hamiltonian density 
 \bb 
 {\cal H}_{\mbox{\tiny{eff}}}= \frac{1}{2}\left( {\bf E}^2 +{\bf B}^2\right) + \xi \left( {\bf E}^2 +{\bf B}^2\right)^2 +\cdots,
 \ee 
 where $\xi$ in this case in the fine structure constant.
  \smallskip
   
The next step we will take is to make an extra deformation of the above non-commutative system. Indeed, 
the system  previously analyzed admits also a different type of algebra deformation. 
That is, instead of considering the relation
as the one in (\ref{conmua}) we posit the relation  \cite{kempf1,kempf2,nos1}
 \bb  
 \left[a_\pm,a_\pm^\dagger \right]= {\cal D} \left[\lambda_\pm a_\pm^\dagger a_\pm \right], 
  \label{sny14}
 \ee 
 where ${\cal D}[x]$ is a deformation operation which, in principle, can be a function of the operator $x$ or an infinite series of $x$ and $\lambda_\pm$ is a dimensionless parameter which, in principle, can be chosen as
 $\lambda_+ =\lambda_- =\lambda$, but we will keep  both different in order to consider a more general scenario.
  
For example if we choose \cite{kempf1} $ {\cal D}[x] = 1-x$ , then (\ref{sny14}) becomes 
 \bb 
 a_\pm a_{\pm}^\dagger -q_\pm~  a_\pm^\dagger a_\pm=1, \label{sny15}
 \ee
 with $q_\pm= 1-\lambda_\pm$ which turn out to be  a representation of the quantum group
$SU_q(2)$ \cite{kulito} for each sector \lq +\rq~ and \lq -\rq.
   
The effect of the deformation is to change the spectrum of 
 $a_\pm^\dag a_\pm$  and  therefore  $\Lambda_{n_-n_+}$  changes. Denoting  by $c_{n_-}(\lambda_-)$ the 
 spectrum of the number operator in the \lq$-$\rq\,sector one gets the spectrum (normal ordered)
 \bb
\Lambda_{n_-n_+} = 2 \theta \, c_{n_-}(\lambda_-),  
\label{sny16}
\ee 
where, for the linear deformation  $ {\cal D}[x] = 1-x$  \cite{kempf1}  one finds  \cite{nos1} 
\begin{equation}
\label{cn}
c_{n}(\lambda) = \lambda^{-1}\left(1-(1-\lambda)^n\right).
\end{equation}

The infinite degeneracy in $n_+$ persists  since it depends on  the fact that $x$ and $y$   in (\ref{sny5}) are 
decoupled sectors. The energy difference between two successive levels for this system turn 
out to be 
\begin{equation}
\Lambda_{ n_{-}+1,n_+} - \Lambda_{n_{-},n_+} =2\theta \,(1-\lambda_-)^{n_-}.
\end{equation}
It  follows that the spectrum is asymmetric, that is, the spectrum is not equally spaced by $ n _- $. 
This  is true for any potential  of the form $ V (\aleph) $. 
%

With this in mind, let us  use the fact that $V$ is a function of $p$ and $x$ (or what is the same, $a$ and 
$a^\dagger$) in order to study different physical systems.  
%
As a concrete example, consider 
\bb 
V(\aleph)= \Omega_N\,\aleph^N, 
\label{potenc} 
\ee
with $\Omega_N$ a constant with dimensions of (energy)$^{2N+1}$.

The  diagonal basis of operator $a^\dag_\pm a_\pm$ is $\{ | n_\pm\rangle \}_{\{n_\pm=0,1,2,\cdots\}}$ and due to the properties 
of the operator $\aleph$, we  consider the base $|n_-,m_+\rangle$, so that the matrix element (we omit the constant 
$\Omega_N$)
$$
 \langle n_- ,n_+|V|m_-,m_+\rangle = \langle n_- |V|m_-\rangle\delta_{n_+,m_+} \equiv  V_{n_-,m_-}\delta_{n_+,m_+} \nonumber
$$
where    the matrix element of the normal ordered operator $V$ is
\eqb 
V_{n_-,m_-} &=& \langle n_-|\aleph. \aleph \cdots \aleph |m_-\rangle \nonumber
\\
&=& \sum_{\{n_i\}} \langle n_-|\aleph |n_1\rangle
\langle n_1|\aleph |n_2\rangle  \cdots
\langle n_N|\aleph |m_-\rangle \nonumber 
\\
&=& \left({2\theta}\right)^N \left(c_{n_-}(\lambda_-)\right )^N 
\delta_{n_-, m_-}. \label{12}
\eqf
  \smallskip
  
 For the particular case of the linear deformation ${\cal D}[x]=1- \lambda x$ with 
  $\lambda\equiv \lambda_- $ we obtain  
\eqb
V_{n_-,m_-}=   \left(\frac{2\theta}{\lambda}\right)^N \left[ 1- (1- \lambda)^{n_-} \right]^N\,\delta_{n_-,m_-}.
\eqf

For the case of Hamiltonian (\ref{sny4}), this result implies
\begin{eqnarray}
\label{hmatrix}
\langle &n_-&,n_+ | H | m_-,m_+\rangle = \langle n_-,n_+ | \frac{{\bf p}^2}{2M} | m_-,m_+\rangle +
\nonumber
\\
&& \Omega_N \left(\frac{2\theta}{\lambda}\right)^N \left[ 1- (1- \lambda)^{n_-} \right]^N\,\delta_{n_-,m_-}\delta_{n_+,m_+}
\end{eqnarray}

However, since we are considering the  lowest  Landau level limit, 
the spectrum (note that  the change of notation $n_- \equiv n$)
$E_{n} \sim V_{n,n}$ becomes  
\bb 
E_{n} =\Omega_N\left(\frac{2\theta}{\lambda}\right)^N \left[ 1- (1- \lambda)^{n} \right]^N.
 \label{turn2}
\ee

Note that this energy is basically the energy of the  {\it turning points} (which further highlights the non perturbative character of this result). 

In order to make contact with the Landau problem described before we take $\Omega_N = \Omega^N$ so that the identification $\Omega\theta = eH_0/2\mu$ holds. Here, $e$ is the electron charge, $\mu$ is a
mass scale and $H_0$ is an external magnetic field. 
\\ 

For $N=1$ (in the limit $M\to\infty$) this model defines  a deformed  Landau problem. The energy can be 
written, alternatively
\begin{eqnarray}
\label{eqsum}
E_n &=& 2\Omega\theta\,\sum_{\ell=0}^{n-1} \frac{n(n-1)\cdots(n-\ell)}{(\ell+1)!}(-\lambda)^\ell
\nonumber
\\
&=&2\Omega\theta\,n\,\left( 1-\frac{n-1}{2!}\lambda\right) + {\cal{O}}(\lambda^2),
\end{eqnarray}
showing that corrections to the energy due to the deformation is order $n^2$. Similar
behavior is observed in the non-relativistic limit of the relativistic Landau problem, however, relativistic
corrections there are also proportional to $(\Omega\theta)^2$.

This suggest that a higher orders of the  $\aleph$ operator might be of physical interest. Consider, for 
example, the potential 
\begin{equation}
\label{ordertwo}
V^{(2)}=\Omega\aleph +\frac{\kappa^2 }4 \Omega^2\aleph^2,
\end{equation}
{{where $\kappa^2$ is a length scale which is determined below by using the quadratic Zeeman effect data and the Breit-Rabi formula. 

According to our previous analysis and also under same assumptions, the energy spectrum turn out}}
to  be
\begin{eqnarray}
E_n^{(2)}&=&\left(\frac{eH_0}{2\mu}\right)  \,n\, \left((2+(1-n)\lambda\right) +
\nonumber
\\
& & \left(\frac{eH_0}{2\mu}\right)^2 \kappa^2\,\,n^2\,\left(1+(1-n)\lambda\right) + {\cal{O}}(\lambda^2). \label{corre33}
\end{eqnarray}

{{ We observe here that linear terms in $\lambda$ has contributions from linear and quadratic terms
of magnetic field with different powers of $n$ and  
\\
\eqb
\Delta E^{(2)}_ n &=& E^{(2)}_ n- E^{(2)}_ {n-1}
\nonumber 
\\
&=&  2 \left(\frac{eH_0}{2\mu}\right) (1+(1-n)\lambda) +
 \nonumber 
\\
&& \kappa^2 \left(\frac{eH_0}{2\mu}\right)^2\,\left[2n-1 +(5n-2-3n^2)\lambda\right]
\eqf

For  non-highly excited levels, the difference of energies becomes  
\bb 
\Delta E^{(2)}_ n \approx2 \left(\frac{eH_0}{2\mu}\right) (1+(1-n)\lambda) +\kappa^2\left(\frac{eH_0}{2\mu}\right)^2 (2 n -1 -3n^2 \lambda). \label{corre}
\ee
}}

In the conventional Landau problem, the difference of frequencies   is a constant  proportional to the external applied magnetic field. In the case we are considering, the
deformation takes an extra factor proportional to $ \kappa^2 \left(\frac{eH_0}{2\mu}\right)^2$ and then, 
if the magnetic field is  large enough,  and using the condition that $\lambda $ can be  properly adjusted, this correction of frequency could be relevant. To be precise, it is enough to have
\begin{equation}
H_0 \gg \left(\frac{2\mu}{e\kappa^2}\right)\frac{(1+(1-n)\lambda)}{(2 n -1 -3n^2 \lambda)}\sim \left(\frac{2\mu}{e\kappa^2}\right)\frac1{3n^2}.
\end{equation}

Actually the $ H_0^2$ contribution due to the deformation plays a similar  role of the quadratic Zeeman effect term. In our case, for strong magnetic field as described before,  one have
\begin{eqnarray}
|\Delta E^{(2)}_ n| &\approx& H_0^2\left(\frac{\kappa e}{2\mu}\right)^2 3n^2\lambda
\\
&\equiv&\kappa_{\mbox{\tiny{eff}}}\,H_0^2
\end{eqnarray}

Although the quadratic corrections of the Zeeman effect are not dominant in atomic physics in general, they are important, for example in the case of alkaline atoms where
\bb
|\Delta \omega |=\kappa_{{\mbox{\tiny{eff}}}} H_0^2.\label{ qzee}
\ee 

These quadratic corrections have been measured in the last twenty  years using different methods and great precision has been achieved with the development of cold atoms 
measurement techniques \cite{rev}. For example for the  $^{87}$Rb ground-state clock transition  $\kappa$ is 
\bb 
\kappa_{{\mbox{\tiny{eff}}}}\sim  575.15\times 10^8 ~{H\mbox{z}~}T^{-2}, \label{experi}
\ee

but from the the Breit-Rabi formula \cite{breit} 
\bb
\Delta \omega = \frac{(g_J-g_I)\mu_B^2}{2 h \Delta H_{Hfs}}H_0^2 = \kappa_{{\mbox{\tiny{eff}}}} H_0^2,  \label{rabi}
\ee 
where $g_{J,I}$ are the Land\'e factors, $\mu_{B}$ the Bohr magneton, $h$ the Planck constant and $\Delta H_{Hfs}$ is the hyperfine energy splitting.
 
In particular the experimental value (\ref{experi}) compared to the $\kappa $ calculated with the Breit-Rabi formula is in excellent agreement with standard measurements \cite{rubi,br}. 

Putting $\kappa$ in natural units then the dimensions of $\kappa$ are
\[
\kappa \sim (\mbox{length})^3,
\]
and replacing (\ref{experi}) we find 
\bb
\kappa^{1/3} \sim ~ \mbox{0.1}~  \r{A}
\ee
which typically could be considered an x-ray regime effect. 

 \section{Conclusions}
 
 {{ In this work we have studied a linear deformation of the Fock algebra and we have applied it to the Landau problem which leads to a family of exactly soluble problems. We consider in detail the quadratic effective Hamiltonian and the length scale is fixed using data from the quadratic Zeeman effect. 
 
 A careful analysis of (\ref{corre}) shows that the validity of our arguments between neighboring levels as predicted by the Breit-Rabi formula \footnote{For a recent experimental  Breit-Rabi formula verification see \cite{br}.}. 
 
 The uncertainties of the Breit-Rabi formula are uncertainties in the physical and atomic constants so that the only possible sources of error may come from the g-factors, but the data considered (cold alkaline atoms \cite{rubi}) seem to be well enough established.}}

%

The case of linear potential, on the other hand, can be solved completely and its extension to the relativistic
case is straightforward and is equivalent to the formulation of non-commutative fields as it was discussed in 
\cite{das}. 
 \\
 
We would like to thank Prof. Jos\'e Luis Cortes for enlightening discussions.  This work was partially supported by Dicyt 041831GR (J.G) and 041931MF (F. M.).


\begin{thebibliography}{99}
\bibitem{Banks:1983by} 
  T.~Banks, L.~Susskind and M.~E.~Peskin,
  Nucl.\ Phys.\ B {\bf 244}, 125 (1984).


  \bibitem{Hawking:1979ig}  R. Penrose, in {\it General Relativity; an Einstein Centenary Survey}, pp. 581, edited by   W. Israel and S. W. Hawking, Univ. Cambridge Press  (1979).
  


  \bibitem{kibble}  T.~W.~B.~Kibble,
  Commun.\ Math.\ Phys.\  {\bf 65}, 189 (1979).  


 
\bibitem{weinberg} S.~Weinberg,
  Annals Phys.\  {\bf 194}, 336 (1989).

  \bibitem{Banks:2020wul} For recents references  see, 
T.~Banks,
[arXiv:2001.07662 [hep-th]]; L.~Amadei, H.~Liu and A.~Perez,
[arXiv:1912.09750 [gr-qc]]; S.~B.~Giddings,
Phil. Trans. Roy. Soc. Lond. A \textbf{377} (2019) no.2161, 20190029; A.~O.~Barvinsky, D.~Carney and P.~C.~E.~Stamp,
Phys. Rev. D \textbf{98} (2018) no.8, 084052.


   \bibitem{polchinski} G.~T.~Horowitz and J.~Polchinski,
  Phys.\ Rev.\ D {\bf 55}, 6189 (1997).  
  
  \bibitem{susskind} See for example, T.~Banks, L.~Susskind and M.~E.~Peskin,
  Nucl.\ Phys.\ B {\bf 244}, 125 (1984).
  

\bibitem{putue}A. Das, J. Gamboa, F. Méndez and F. Torres, Phys. Lett. A {\bf 375}, 1756 (2011).



\bibitem{snyder} H.~S.~Snyder,
  Phys.\ Rev.\  {\bf 71}, 38 (1947).
  


   \bibitem{yang} 
  C.~N.~Yang,
  Phys.\ Rev.\  {\bf 72}, 874 (1947).  

\bibitem{saco} T.~Curtright, T.~Uematsu and C.~K.~Zachos,
J. Math. Phys. \textbf{42} (2001), 2396; T.~L.~Curtright and C.~K.~Zachos,
Asia Pac. Phys. Newslett. \textbf{1} (2012), 37-46; T.~Curtright, D.~Fairlie and C.~K.~Zachos,
Phys. Rev. D \textbf{58} (1998), 025002

  
  \bibitem{connes} A. Connes, {\it Noncommutative Geometry},  Academic Press, 1994.  



  \bibitem{kulito} Drinfel'd, V.G. 
J Math Sci {\bf 41}, 898 (1988);  M.~Chaichian and P.~Kulish,
  Phys.\ Lett.\ B {\bf 234}, 72 (1990); A.~J.~Macfarlane,
  J.\ Phys.\ A {\bf 22}, 4581 (1989); L.~C.~Biedenharn,
  J.\ Phys.\ A {\bf 22}, L873 (1989).

  \bibitem{konse} M.~Kontsevich,
  Lett.\ Math.\ Phys.\  {\bf 66}, 157 (2003);  ibid.   M.~Kontsevich,
  Commun.\ Math.\ Phys.\  {\bf 147}, 1 (1992); F.~Bayen, M.~Flato, C.~Fronsdal, A.~Lichnerowicz and D.~Sternheimer,
  Annals Phys.\  {\bf 111}, 61 (1978); A.~S.~Cattaneo and G.~Felder,
  Commun.\ Math.\ Phys.\  {\bf 212}, 591 (2000).



\bibitem{nos01} J.~Gamboa, M.~Loewe, F.~Mendez and J.~C.~Rojas,
  Int.\ J.\ Mod.\ Phys.\ A {\bf 17}, 2555 (2002); ibid, J.~Gamboa, M.~Loewe, F.~Mendez and J.~C.~Rojas,
  Mod.\ Phys.\ Lett.\ A {\bf 16}, 2075 (2001).
  \bibitem{refers} This is a topic extensively studied in the last twenty years and here we follow our work and other ones see; 
  L.~Mezincescu,
[arXiv:hep-th/0007046 [hep-th]]; 
 J. Gamboa, M. Loewe and J.C.Rojas, Phys.\ Rev.\ D {\bf 64}, 067901 (2001); 
V.~Nair and A.~Polychronakos,
Phys. Lett. B \textbf{505}, 267-274 (2001);
C. Duval  and P. A. Horvathy,  Phys. Lett. B {\bf 479}, 284 (2000);
C. Duval  and P. A. Horvathy, J. Phys. A: Math. Gen.  {\bf 34}, 10097 (2001).



  \bibitem{gross} See for example,  D.~J.~Gross, J.~Kruthoff, A.~Rolph and E.~Shaghoulian,
  Phys.\ Rev.\ D {\bf 101}, no. 2, 026011 (2020) and references therein.  


  \bibitem{kempf1} A.~Kempf,
  J.\ Math.\ Phys.\  {\bf 35}, 4483 (1994.


  \bibitem{kempf2}  A.~Kempf,
  J.\ Phys.\ A {\bf 30}, 2093 (1997).
  
    \bibitem{nos1} J.~L.~Cort\'es and J.~Gamboa,``Deformed classical-quantum mechanics transition,''  , arXiv:2004.05673 [hep-th].  





  

\bibitem{breit} G. ~Breit and I. ~I. ~Rabi
\, Phys. \, Rev. \, {\bf 38}, 2082 (1931).  


\bibitem{rev} For references and a review see for example, 
B-~M.~ Garraway and H.~ Perrin, J. Phys. {\bf 49} B, 172902 (2016).
\bibitem{rubi} D.~A.~Steck, 2010 Rubidium 87 D Line Data, http:7steck.us/alkalidata
 

\bibitem{br} B.  Wu et.~ al. ~J.\  Opt.\ Soc. \ Am. ~B {\bf  31}, 742 (2014).


  

\bibitem{das} J.~M.~Carmona, J.~L.~Cortes, A.~K.~Das, J.~Gamboa and F.~Mendez,
Mod. Phys. Lett. A \textbf{21} (2006), 883-892%
%
%
%
%
%
%
%
%
%
  
  \end{thebibliography}
\end{document}